\documentclass[11pt]{article}

\usepackage{amsfonts}
\usepackage{amssymb}
\usepackage{amsmath}
\usepackage{amsthm}
\usepackage{amscd}
\usepackage{graphicx}
\usepackage{color}
\usepackage{bbm}
\usepackage{comment}
\usepackage{mathrsfs}
\usepackage{hyperref}
\usepackage{lipsum}
\usepackage{subcaption}
\usepackage{enumitem}

\usepackage{tikz}
\usetikzlibrary{quantikz}

\title{Exploring the optimality of approximate state preparation quantum circuits with a genetic algorithm}

\author{{\bf Tom Rindell$^{1,2}$, Berat Yenilen$^{3}$, Niklas Halonen$^{1,}$,}\\{\bf Arttu Pönni$^{1,}$, Ilkka Tittonen$^1$, Matti Raasakka$^{1,}$\thanks{Corresponding author, email: \tt{matti.raasakka@aalto.fi}}}\vspace{5pt}\\\small $^1$Micro and Quantum Systems group, Dept. of Electronics and Nanoengineering, Aalto University\\\small $^2$Department of Physics, University of Helsinki\\\small $^3$Department of Physics, RWTH Aachen University}

\date{\today}

\begin{document}

\maketitle

\begin{abstract}
	We study the approximate state preparation problem on noisy intermediate-scale quantum (NISQ) computers by applying a genetic algorithm to generate quantum circuits for state preparation. The algorithm can account for the specific characteristics of the physical machine in the evaluation of circuits, such as the native gate set and qubit connectivity. We use our genetic algorithm to optimize the circuits provided by the low-rank state preparation algorithm introduced by Araujo et al., and find substantial improvements to the fidelity in preparing Haar random states with a limited number of CNOT gates. Moreover, we observe that already for a 5-qubit quantum processor with limited qubit connectivity and significant noise levels (IBM Falcon 5T), the maximal fidelity for Haar random states is achieved by a short approximate state preparation circuit instead of the exact preparation circuit. We also present a theoretical analysis of approximate state preparation circuit complexity to motivate our findings. Our genetic algorithm for quantum circuit discovery is freely available at \href{https://github.com/beratyenilen/qc-ga}{\tt{https://github.com/beratyenilen/qc-ga}}.
\end{abstract}

\section{Introduction}
Quantum computing has gained a lot of attention in the recent years mainly due to the continued development of quantum hardware by both academic research groups and companies. Significant speed-ups in certain computational tasks by quantum computers were theoretically demonstrated already in the 1990's by several authors \cite{Deutsch92,Simon97,Shor,Grover}. However, a quantum advantage for specific computational problems was only recently experimentally demonstrated \cite{Google19,Zhong20}. This has brought the field past the proof-of-concept stage to a point, where real-world applications for quantum computing are starting to emerge, although still severely limited by the hardware. First applications are expected to be simulations of quantum systems and quantum machine learning, as these typically have lower hardware requirements compared to many other potential applications, which require higher computational fidelity and thus quantum error correction with additional overhead \cite{Preskill18}.

The application of genetic algorithms to quantum circuit discovery has already been discussed in the literature for quite a while (see e.g. \cite{Spector98,Williams99,Lukac02,Spector04,Massey06,Ruican07}). Automating the design of quantum algorithms is an appealing idea, as the behavior of quantum systems, such as a quantum processor, is often unintuitive and challenging to understand. However, only recently the growth of classical and quantum computational resources has started to allow for the practical feasibility of this approach. Genetic algorithms have recently been succesfully applied to quantum circuit optimization, e.g., in quantum machine learning \cite{AltaresLopez21,Lu21,Arufe22} and chemistry \cite{Chivilikhin20}, and specialized software packages have been developed for the task \cite{Sunkel23}. The original inspiration for our work was however the paper \cite{Potocek} by Poto\v cek et al., who obtained promising results in applying a multi-objective genetic algorithm to discover optimized quantum circuits for the implementation of quantum Fourier transform and Grover search. With the multi-objective approach not only one finds circuits that reproduce the exact desired unitary, but the algorithm can also produce a set of alternative circuits with variable trade-offs between accuracy and circuit depth. This can be highly desirable, when running the circuit on physical hardware, as the error rate grows exponentially as a function of the circuit depth in the absence of error correction.

State preparation is among the most fundamental tasks in quantum computing. Indeed, essentially all quantum algorithms proceed by preparing some quantum state for the qubits, whose statistics contains information for solving the computational problem at hand. The state preparation problem can be stated as follows: Given some set of basis gates, find a quantum circuit composed of these gates, which transforms the all-zeros state $|00\ldots0\rangle$ to some target state $|\psi\rangle$ \cite{Shende06,Plesch11}. Besides being a suitable first problem to gauge the performance of our genetic algorithm for circuit discovery, this problem also has practical significance. If the same state needs to be prepared a large number of times during the running of the algorithm, as is often the case due to the probabilistic nature of quantum computing, small gains in the preparation process of the state may lead to significant savings in the running time of the algorithm.

The minimum number of gates required to reach a certain state (possibly with a margin of error) from a fixed reference state is called the quantum circuit complexity of a quantum state \cite{Yao93,Watrous2008}. Our genetic algorithm for state preparation provides estimates of the quantum circuit complexity of multi-qubit states. On the theory side, quantum circuit complexity is of fundamental importance for several reasons. Most fundamentally, the difficulty of a computational task can be quantified by how the circuit complexity scales as a function of the input size. Understanding the quantum circuit complexity of different computational problems has far-reaching practical consequences, e.g., for the safety of cryptographic protocols. In condensed matter theory, phase transitions have recently been studied with quantum circuit complexity \cite{Liu20,Xiong20,Roca-Jerat23}. On the other hand, in quantum gravity circuit complexity has been the topic of intense research, in particular in relation to the anti de Sitter/conformal field theory (AdS/CFT) duality through the `complexity$=$volume' \cite{Susskind14,Couch18} and `complexity$=$action' \cite{Brown16} conjectures, with attempts to relate quantum circuit complexity of quantum field theory states to spacetime geometry.

In this work, we developed a genetic algorithm enabling us to find quantum circuits, which provide trade-offs between weighted circuit depth (`circuit cost') and the fidelity of the output state. We then applied the genetic algorithm to find approximate state preparation circuits for 1000 Haar random states of 5 qubits, using the qubit connectivity and the noise model of a 5-qubit IBM Falcon 5T processor provided by IBM Qiskit. More specifically, we first applied the low-rank state preparation (LRSP) algorithm introduced in \cite{Araujo21} to the states, and then further optimized the circuits with the genetic algorithm to study their optimality. For all the states, the fidelity with noise is maximized by an approximate circuit. Theoretical analysis shows that, in the absence of noise, the fidelity of approximate circuits approaches unity exponentially as the circuit depth increases, while noise causes an exponential decay in fidelity as a function of the circuit depth. It is the delicate balance of these two effects, which allows to find approximate circuits that maximize the noisy fidelity. We find that the maximal noisy fidelity of the output states is increased by 9.4\% on average with the circuits found by the genetic algorithm as compared to the circuits given by the LRSP algorithm, with maximum observed relative improvement of 38\%.

The structure of the rest of the paper is as follows. In Section \ref{sec:stateprep} we discuss the state preparation problem in more detail. We also discuss the effect of noise on the state preparation problem on physical hardware, and provide a theoretical analysis of the relation between circuit depth and state preparation fidelity. Section \ref{sec:genalg} describes the detailed implementation of our genetic algorithm. In Section \ref{sec:results} we explore the performance of our algorithm, and show that we improve on the LRSP algorithm results. Section \ref{sec:summary} provides first a summary and discussion of the results. Finally, we provide some pointers for the direction of future work.

\section{State preparation problem in the presence of noise} \label{sec:stateprep}

\subsection{Problem description}
Let us restate what we call in this paper the \emph{exact state preparation problem} more precisely. Let $n$ be a positive integer. We are given
\begin{itemize}
	\item an $n$-qubit state $|\text{target}\rangle$, the \emph{target state}, and
	\item a universal set of quantum gates $G$.
\end{itemize}
Quantum gates are typically assumed to be simple unitaries, which act only on a few qubits at a time.\footnote{Some implementations of quantum processors, such as ion traps, may allow for gates, which act on all the qubits simultaneously. However, here we mostly concentrate on superconducting devices, such as transmon qubits.} Some gates, such as a 1-qubit rotation gate $R_z(\theta)$, may be parametrized by one or more real parameters. Universality of the gate set implies that we can approximate any $n$-qubit unitary to an arbitrary precision $\epsilon > 0$ by a finite sequence of gates from the gate set. Several examples of universal gate sets exist. However, in this work we mainly concentrate on the universal gate set
\begin{align*}
	G_\text{IBM} = \{R_z, \sqrt{X}, X, \text{CNOT}\} \,,
\end{align*}
which has been experimentally implemented as the native gate set of IBM quantum processors. The exact state preparation problem then asks us to find a sequence of gates $g_lg_{l-1}\cdots g_2g_1$, $g_k\in G\ \forall k$, such that
\begin{align*}
	g_lg_{l-1}\cdots g_2g_1 |00\ldots 0\rangle = |\text{target}\rangle \,.
\end{align*}
Upper bounds on the number of CNOT gates for exact state preparation without ancilla qubits have been derived and improved upon by several different authors \cite{Shende06,Mottonen05,Bergholm05,Plesch11,Iten16}.\footnote{In this paper we only consider the case without ancilla qubits. With the use of ancilla qubits the circuit depth can be significantly reduced, see e.g. \cite{Zhang21,Zhang22}. The closely related task of unitary synthesis has also been the subject of active research efforts recently, see e.g. \cite{Iten16,Davis19,Camps20,Younis20}.} Typically only the CNOT count is considered for near-term applications, as the error rates of 1-qubit gates are an order of magnitude smaller. Often the connectivity of the qubits is not restricted in deriving these results, which means that the CNOT gates may act on any pair of qubits. The strictest bounds to our knowledge on the number of CNOT gates for all-to-all connectivity are given by \cite{Plesch11,Iten16}:
\begin{align*}
	C_\text{ub} = \left\{ \begin{array}{ll} 	\frac{23}{24} 2^n - \frac{3}{2}2^{\frac{n+1}{2}} + \frac{4}{3}, & n \text{ odd}\ (n\geq 5)\\
		\frac{23}{24} 2^n - 2^{\frac{n}{2}+1} + \frac{5}{3}, & n \text{ even} \end{array} \right. \,.
\end{align*}
However, the qubit connectivity may also be restricted without changing the problem significantly. This brings the problem closer to practice, as the connectivity of qubits is usually heavily restricted on physical hardware, but makes theoretical analysis more challenging. The only work we found considering restricted connectivity is \cite{Bergholm05}, where the authors consider the state preparation problem on a linear chain of qubits. The difference in CNOT counts between different topologies is always polynomial, as one may permute qubits using a polynomial number of SWAP gates, but it may still have a significant contribution in practice.

Also, the approximate state preparation problem may be considered, which asks us to find a sequence of gates $g_lg_{l-1}\cdots g_2g_1$, $g_k\in G\ \forall k$, such that
\begin{align*}
	|\langle\text{target} | g_lg_{l-1}\cdots g_2g_1 |00\ldots 0\rangle |^2 < \delta
\end{align*}
for some fixed $\delta > 0$. Upper bounds on the number of CNOT gates in these cases are much less known. A recent approach to numerically estimate the bound for a small number of qubits is found in \cite{Ashhab22}. However, we have been unable to find a theoretical analysis of the approximate state preparation problem. In the next subsection (and in \ref{app:error}) we provide a back-of-the-envelope theoretical analysis of the optimal trade-off between fidelity and CNOT count, which we later confirm with our numerical results obtained with the genetic algorithm.

\begin{figure}
	\centering\includegraphics[width=0.75\textwidth]{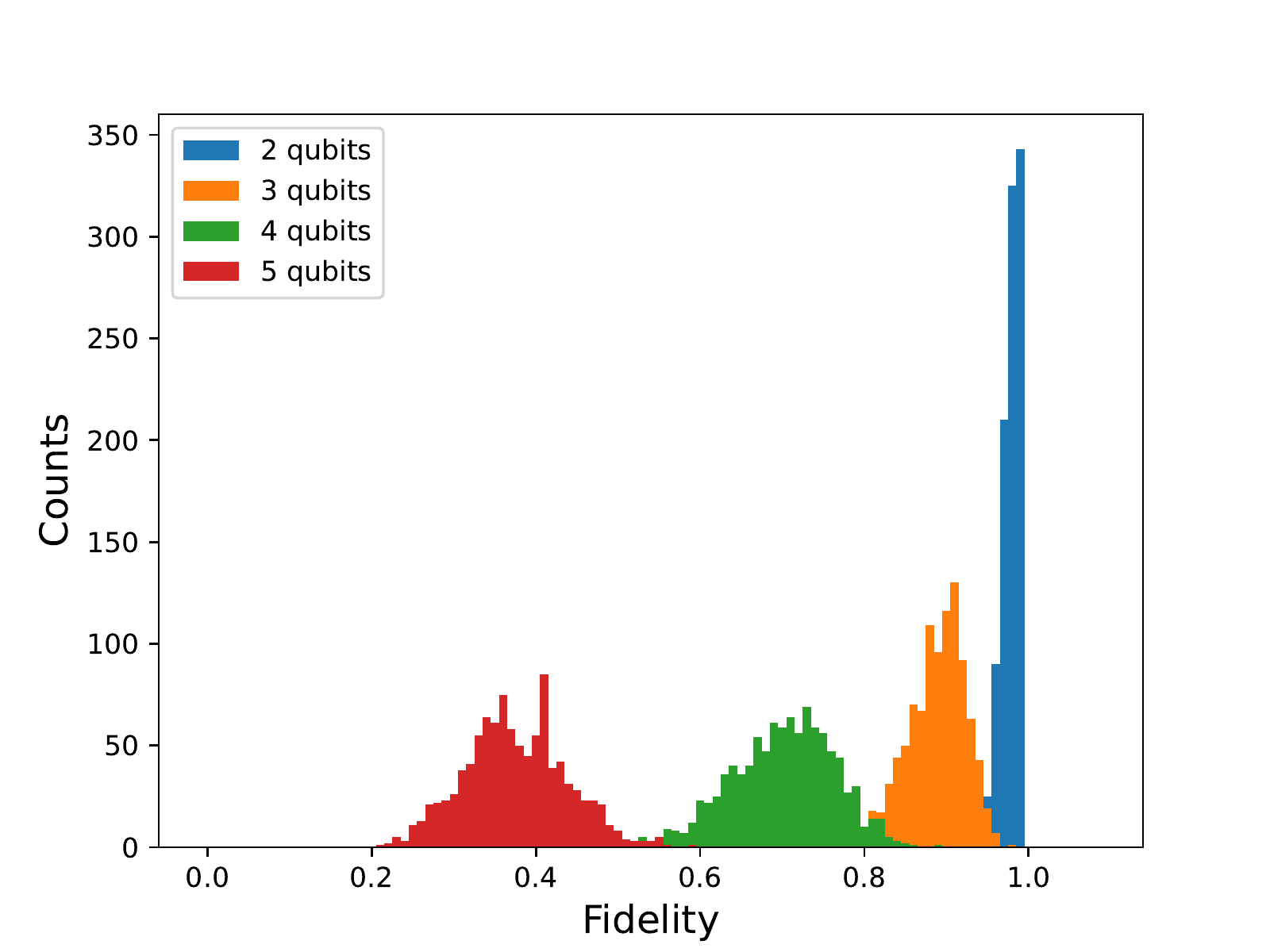}
	\caption{Histogram of noisy fidelities for 1000 simulated state preparation circuits for 2, 3, 4 and 5 qubits provided by the standard Qiskit library method. The simulations were performed using the IBM Qiskit FakeVigo backend.}
	\label{fig:qiskitfids}
\end{figure}

The state preparation problem for a physical quantum computer is somewhat different from the abstract version of the problem due to the presence of noise. Even if we could find the absolute minimum depth circuit, which prepares the target state exactly, this may still not be the best circuit in practice. Producing a generic state of $n$ qubits exactly requires an exponential $\sim 2^n$ number of gates, which makes the circuit basically useless with the non-negligible levels of noise in the current hardware, for just a moderate number of qubits. This is demonstrated in Fig.~\ref{fig:qiskitfids}, where we have plotted the performance of the standard IBM Qiskit exact state preparation method (based on the algorithm presented in \cite{Shende06}) for an increasing number of qubits in the presence of simulated noise. Therefore, what we ideally want to find is a circuit, which offers an optimal trade-off between the circuit depth (or the CNOT count) and the noise-free error in the target state such that the total error in the presence of noise is minimized. This is where a genetic algorithm becomes particularly useful, as it can perform simultaneous multi-objective optimization on these parameters. In Section \ref{sec:genalg} we will explore in detail the application of our genetic algorithm to the state preparation problem.

\subsection{Theoretical analysis}\label{subsec:theory}
On a theoretical level, the trade-off between circuit length and fidelity can be understood in the following way. First of all, let us consider the dependence of the final state fidelity on the length of the circuit in the presence of noise. For simplicity imagine that all our gates have the same error rate $p$. Then the probability of no errors for a circuit of length $l$ is $P(\text{no error}) = (1-p)^l$. Assuming that any error takes us to an approximately orthogonal state, we get for the final state fidelity
\begin{align*}
	F(\rho_\text{final}, U_\text{circ}|00\ldots 0\rangle ) \approx (1-p)^l \,,
\end{align*}
which shows an exponential decrease in fidelity as the circuit length increases. In practice, at least for a superconducting NISQ device, we may consider $l$ to be the number of CNOT gates in the circuit to get a reasonable estimate, since the single-qubit gates have error rates that are typically an order of magnitude smaller.

On the other hand, we may reduce the effects of noise by employing a shorter circuit, which approximates the target state. In \ref{app:error} we estimate the dependence of the noise-free error $\epsilon$ on the CNOT count $l$ in preparing an arbitrary state. We find that (given certain assumptions discussed in more detail in the appendix) we may approximate
\begin{align*}
	\epsilon \approx \exp\left[ - \frac{A(n)}{2} \frac{l}{l_\text{ph}} - \frac{B(n)}{2} \right] \,,
\end{align*}
where
\begin{align*}
	A(n) = \frac{(2 \ln 2)n + \ln\left( \frac{n(n-1)}{2} \right)}{2^n-n-1} \,,\quad B(n) = \frac{(\ln 2)n^2}{2^n-n-1} \,,
\end{align*}
and $n$ is the number of qubits. Moreover, $l_\text{ph}$ is the average number of physical CNOT gates required to implement one logical CNOT gate, which reflects the connectivity of the qubits. We observe that the error $\epsilon$ decreases exponentially as a function of the number of CNOT gates $l$. This is essentially due to the fact that the number of states or, more precisely, the fraction of the Hilbert space volume reached with a fixed error $\epsilon \ll 1$ increases exponentially in $l$: Each CNOT gate allows for two more single-qubit rotations (each with two independent parameters) to be added after it, thus increasing the dimensionality of the subspace of states, which can be reached without error, by 4. Since the dimensionality grows linearly in $l$, the volume grows exponentially. Thus, the error $\epsilon$ needed to cover the whole Hilbert space (i.e., to reach the volume of the full Hilbert space) decreases exponentially in $l$.  Accordingly, we should be able to get a decent approximation to any state with a relatively small number of CNOT gates as compared to the circuit producing the exact state.

Now, imagine we want to produce the target state $|\text{target}\rangle$ with our noisy quantum computer. We may reach noise-free fidelity
\begin{align*}
	F(|\text{approx}\rangle, |\text{target}\rangle ) = 1 - \epsilon^2  \approx 1 - \exp\left[  - A(n) \frac{l}{l_\text{ph}} - B(n) \right]
\end{align*}
with a circuit containing at most $l$ CNOT gates and outputting the approximation state $|\text{approx}\rangle$. When we run this circuit on the noisy quantum computer, the fidelity reduces approximately by the factor $(1-p)^l$, where $p$ is the error rate of the CNOT gates.\footnote{As we later note in fitting these formulae to our numerical results, the error rate $p$ also accounts implicitly for 1-qubit gates errors, decoherence and other possible errors that crop up during the computation. Thus, we may expect $p$ to be larger in practice than the bare CNOT error rate.} Thus, we reach a total fidelity
\begin{align*}
	\quad F(\rho_\text{noisy}, |\text{target}\rangle ) \approx (1-p)^l \left(1 - \exp\left[  - A(n) \frac{l}{l_\text{ph}} - B(n) \right] \right) \,.
\end{align*}
The total fidelity in the presence of noise has a global maximum at
\begin{align*}
	\frac{l_*}{l_\text{ph}}  = \frac{1}{A(n)} \ln\left( 1 - \frac{A(n)}{\ln(1-p)} \right) - \frac{B(n)}{A(n)}\,,
\end{align*}
meaning that at length $l_*$ the trade-off between circuit accuracy and noise is optimized. (See \ref{app:asymptotics} for an asymptotic analysis of these formulae.)

Unfortunately, in practice the number $l_\text{ph}$ of physical CNOT gates per one logical CNOT gate is not easy to compute exactly, since the distribution is unknown. However, a lower bound is provided by the average distance $\langle d \rangle$ between qubit pairs, $l_\text{ph} \geq \langle d \rangle$, since in order to implement a CNOT between an arbitrary pair of qubits we need to at least connect the qubits via CNOT gates. An upper limit is provided by $l_\text{ph} \leq 1 + 6(\langle d \rangle - 1)$, since we may always move distant qubits next to each other by permuting qubits with SWAP gates before applying the CNOT gate, and then move the qubits back to their original positions afterwards. (However, this is usually not the optimal way to perform the logical CNOT gate.)  Each SWAP gate requires 3 CNOT gates, so altogether we need 6 CNOT gates to move a qubit one step forward and then back. Using these estimates, we find that for a small number of qubits a significant reduction of noise can be achieved by using short approximate circuits for state preparation tasks. For example, for a linear chain of 10 qubits ($n=10$, $l_\text{ph} \sim \langle d \rangle \approx 3.7$) with a CNOT error rate $p=0.01$ we find the maximum value of $F(\rho_\text{noise}, |\text{target}\rangle )_\text{max} \approx 0.163$ at circuit length $l_* = 67$. For the exact state preparation circuit of length $l\approx 2^{10}l_\text{ph}$ the total fidelity drops essentially down to zero. In Section \ref{sec:results} we will see that these formulae are reproduced reasonably well by our numerical experiments.

\section{Multi-objective optimization using genetic algorithms}\label{sec:genalg}
Genetic algorithms are heuristic optimization algorithms motivated by the evolutionary processes in nature \cite{Spector04,Konak06}. The basic idea is that we have a population of candidate solutions to the problem described by their genomes, which are iteratively refined though selective breeding. Individual solutions are selected from the population for breeding (i.e., to mutate or combine with other selected solutions) based on their fitness, which measures how well they solve the desired problem. Diversity in the population is maintained by mutations. The selection pressure on the population guides it towards more optimal solutions. Genetic algorithms seem to be particularly well suited to global optimization problems, which present a complex fitness landscape.

Since genetic algorithms deal with populations of candidate solutions, they are also naturally adapted to multi-objective optimization problems \cite{Konak06}. In this case there exists several different properties of the solutions, which we try to optimize simultaneously, but which cannot be simultaneously optimized to their global minima/maxima. In this case, the best we can do is to find solutions with a reasonable trade-off between the different desired properties. Several multi-objective genetic algorithms have been developed in the past, such as Fast Non-dominated Sorting Genetic Algorithm (NSGA2) and Improved Strength Pareto Evolutionary Algorithm (SPEA2), which perform very well against many different kind of problems. In this work we apply the NSGA2 algorithm \cite{nsga2} for ranking and selection of the candidate solutions, as it is implemented in the DEAP Python package \cite{DEAP}, while genetic representation and the evolutionary operations of mutation and cross-over for quantum circuits are adapted from the work of Poto\v cek et al. \cite{Potocek}. Let us now describe some of the details of the implementation. Our full implementation can be found at \cite{code}.

The problem instance is defined by the target state $|\text{target}\rangle$ and the set of allowed gates. In this work, we focus on the IBM quantum processors with the native gate set
\begin{align*}
	G_\text{IBM} = \{ R_z(\theta), X, \sqrt{X}, \text{CNOT} \}\,
\end{align*}
where $R_z(\theta) = e^{iZ\theta/2}$, $\theta\in [0,2\pi)$ is the parametrized rotation by angle $\theta$ around the $z$-axis on the Bloch sphere. This set of gates constitutes a universal gate set, where the only native multi-qubit gate is the CNOT gate. Moreover, the allowed CNOT gates are limited by the qubit connectivity of the quantum processor. A quantum circuit $C$ is represented in the genetic algorithm by a genome, which is just a list of tuples of form $(g,q)$, where $g\in G_\text{IBM}$ and $q$ is a list of indices specifying the qubits on which the gate $g$ acts. The operations, which are used to modify circuits, are defined as follows:
\begin{itemize}
	\item Operations acting on single gates within one circuit:
	\begin{enumerate}
		\item {\bf Discrete uniform mutation:} Iterate over all gates in the circuit and randomly change the target and/or control qubits with probability inversely proportional to the length of the circuit.
		\item {\bf Continuous uniform mutation:} Iterate over all gates in the circuit and adjust the parameter for the $R_z(\theta)$ gates.
		\item {\bf Move gate:} A randomly chosen gate is moved to a new randomly chosen position in the circuit.
		\item {\bf Insert mutate invert:} Apply a discrete mutation on a single gate, after which a randomly selected gate and its inverse are placed immediately before and immediately after, respectively.
	\end{enumerate}
	\item Operations acting on sequences of gates within one circuit:
	\begin{enumerate}[resume]
		\item {\bf Sequence insertion:} Generate a random gate sequence, which is then inserted at a random point in the gate list.
		\item {\bf Sequence and inverse insertion:} Apply a sequence insertion at one point and insert the inverse sequence at a later point in the gate list.
		\item {\bf Sequence deletion:} Delete a randomly selected interval in the gate list.
		\item {\bf Sequence replacement:} Sequence deletion followed by sequence insertion.
		\item {\bf Sequence swap:} Swap two randomly chosen sequences in the gate list.
		\item {\bf Sequence scramble:} Pick a sequence in the gate list randomly, and perform a random permutation on that sequence.
	\end{enumerate}
	\item Operations acting on two circuits:
	\begin{enumerate}[resume]
		\item {\bf Crossover:} Select a random number of gates from the beginning of the first parent circuit, disregard the same number of gates from the second parent, and combine them into a child circuit. Additionally, a second child with the opposite combination is returned as well.
	\end{enumerate}
	\item Operations modifying other properties of a single circuit:
	\begin{enumerate}[resume]
		\item {\bf Permutation mutation:} Permute the mapping of logical qubits to physical qubits.
		\item {\bf Clean:} Remove all pairs of subsequent gates that are inverses of each other, and combine subsequent pairs of $z$-rotations. Iterate until no simplifications are found.
	\end{enumerate}
\end{itemize}
Single-gate operations 1--4 depend on the expected mutation count (EMC) parameter, which is a fixed constant during the evolution. For these operations the probability of a mutation on a single gate is given by $\text{EMC}/l$, where $l$ is the length of the gate list. Operation 2 also takes an additional continuous mutation width (CMW) parameter for the adjustment of parametrized gates. The gate parameters are adjusted by adding a value from Gaussian distribution with standard deviation $\text{CMW}/\epsilon$, where $\epsilon$ the error of the circuit. With this approach the parameter will be adjusted less for more optimal circuits. Sequence operations 5--10 pick the length of the sequence randomly from a geometric distribution with the mean value given by the expected sequence length (ESL) parameter, which is another fixed constant. Constants EMC, CMW and ESL are all set to 2.0 by default, following Poto\v cek et al. \cite{Potocek}. While the evolutionary operations above were mostly adapted from Poto\v cek et al. \cite{Potocek}, some have been altered for the purposes of this implementation.

Before running the genetic algorithm, parameters such as the population size, the number of generations, the number of qubits, and the simulation backend must be initialized. Some parameters, such as the noise model, basis gates, and the connectivity can be taken, e.g., from the desired simulated Qiskit backend, but can also be adjusted if needed. The population is initialized with circuits obtained from the LRSP algorithm \cite{Araujo21}. The remaining population is filled with randomly generated circuits. The size of the population stays constant during the evolution.

Candidate circuits in every generation are evaluated with respect to two criteria:
\begin{enumerate}
	\item {\bf Circuit cost:} Defined as
	\begin{align*}
		\text{cost} = \text{(number of single-qubit gates)} + 10 \times \text{(number of CNOT gates)} \,.
	\end{align*}
	\item {\bf Fidelity:} Apply the unitary $U_C$ corresponding to the circuit $C$ to the initial state $|00\ldots 0\rangle$, and compute the fidelity
	\begin{align*}
		F(|\text{target}\rangle,U_C|00\ldots 0\rangle) = |\langle\text{target}|U_C|00\ldots 0\rangle |^2
	\end{align*}
	with respect to the target state.\footnote{In the current implementation the application of the unitary $U_C$ and the computation of the fidelity is done via classical simulation. This is the most resource-intensive part of the algorithm. However, one could imagine moving this part of the algorithm to be performed by a quantum computer if one had an access to a good enough machine.}
\end{enumerate}
In comparing circuits, a circuit is said to dominate those circuits, which have both a lower fidelity and a higher cost. The optimal `non-dominated' circuits are those, which are not dominated by any other circuit. After evaluation the circuits are sorted in a list from best to worst using the multi-objective NSGA-2 non-dominated sorting algorithm. Essentially, the circuits are ranked according to their dominance: Non-dominated circuits get rank 1; those circuits, which are dominated only by rank 1 circuits, get rank 2, etc.

The production of the next generation proceeds as follows. First, the best 10\% of the population according to their rank are moved directly into the next generation. If the circuits from a certain rank do not all fit into the 10\%, the chosen ones are picked randomly. Then, the remaining 90\% of the next generation is filled with modified versions, i.e., offspring of the circuits in the current generation. First, the rank of the parent circuit is picked according to a Boltzmann-like distribution, where the probability of a rank $r$ to be chosen is proportional to $e^{-r}$. After the rank has been chosen, any circuit is picked from that rank with an equal probability. New candidates are generated from the chosen circuits with evolutionary operations listed above until the population is filled for the next generation. Evolutionary operations are picked randomly with equal probability. This procedure results in a new generation of circuits with an altered population. The process is iterated over for a given number of generations.

Once the algorithm has been run for a given number of generations, a final `clean' operation is performed on every circuit of the final population to remove all the gates that trivially cancel each other. The final population is then saved alongside a DEAP logbook file that contains statistical data for every generation.

\section{Results}\label{sec:results}
For our target states we generated 1000 random statevectors for 5 qubits, sampled from the uniform Haar measure. For each state, the genetic algorithm was seeded with circuits obtained from the LRSP algorithm (transpiled using the Qiskit transpiler), and then run on the Triton supercomputer cluster \cite{Triton} with population of 400 for 30 000 generations. Without multithreading, the duration of a single run for one state varied from 32 to 50 hours. The backend simulation and noise model were implemented from IBM Qiskit's `FakeVigo' backend, which models a 5-qubit IBM Falcon 5T processor.

\begin{figure}[h!]
	\includegraphics[width=0.5\textwidth]{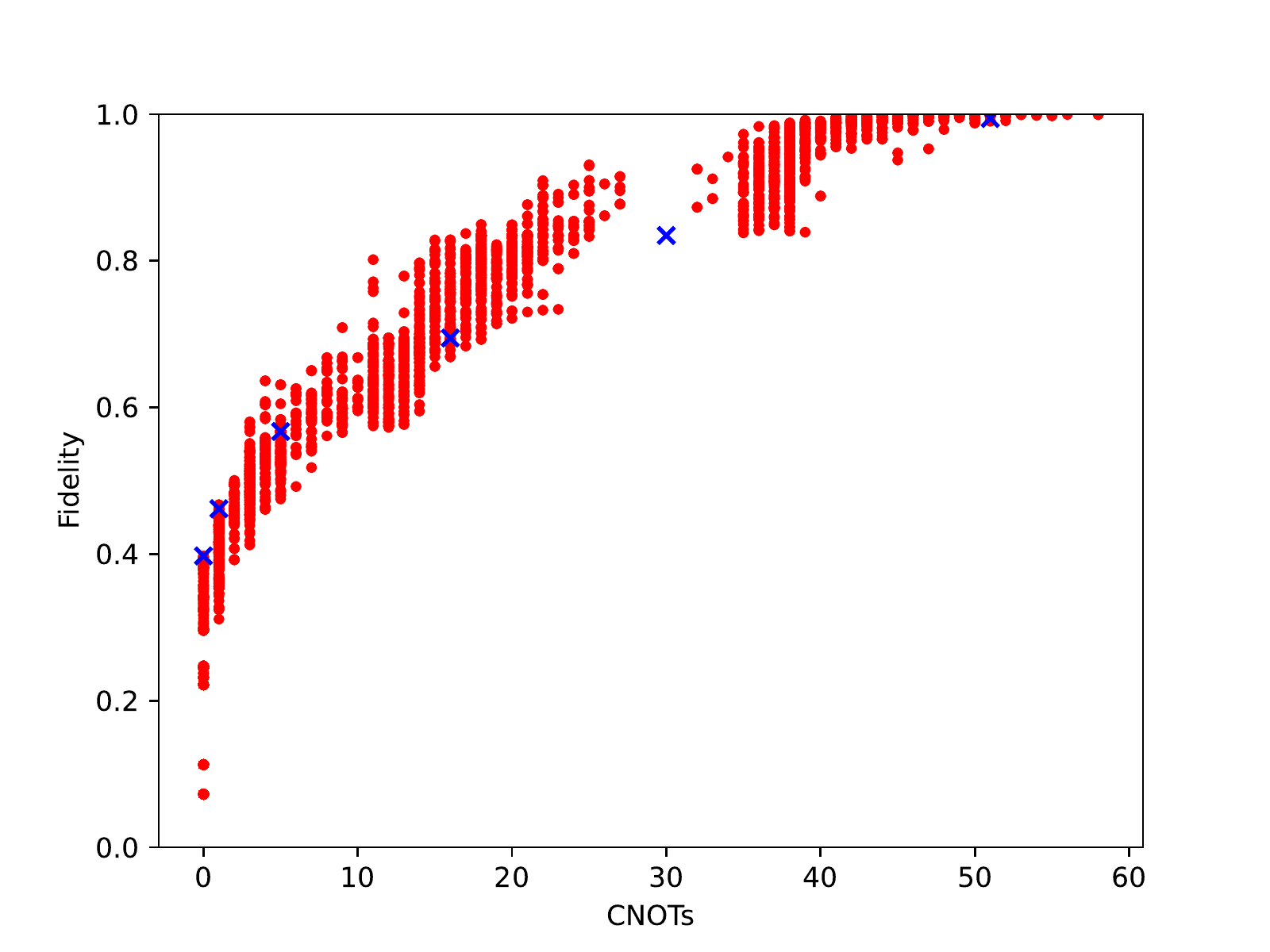}
	\includegraphics[width=0.5\textwidth]{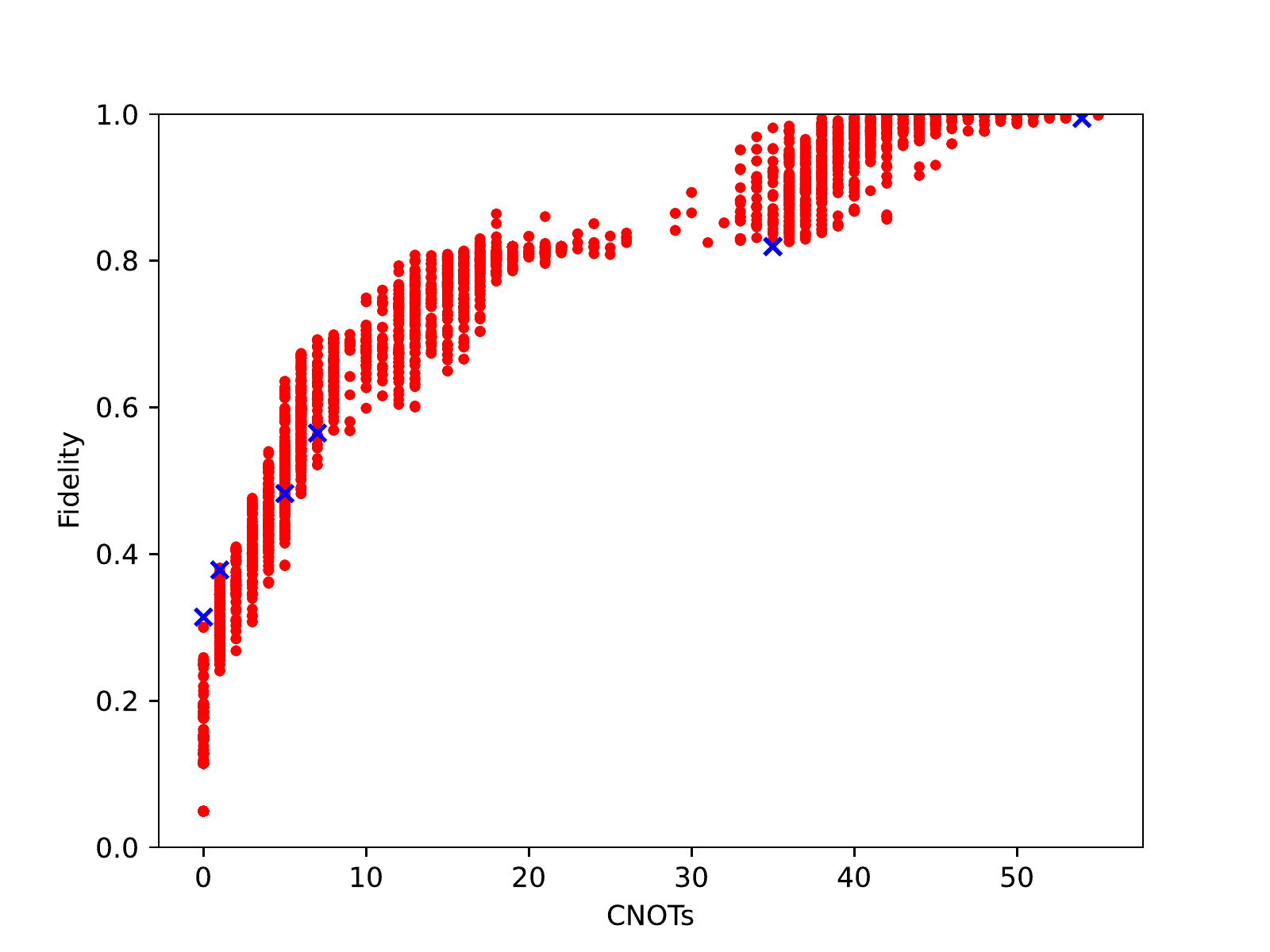}
	\includegraphics[width=0.5\textwidth]{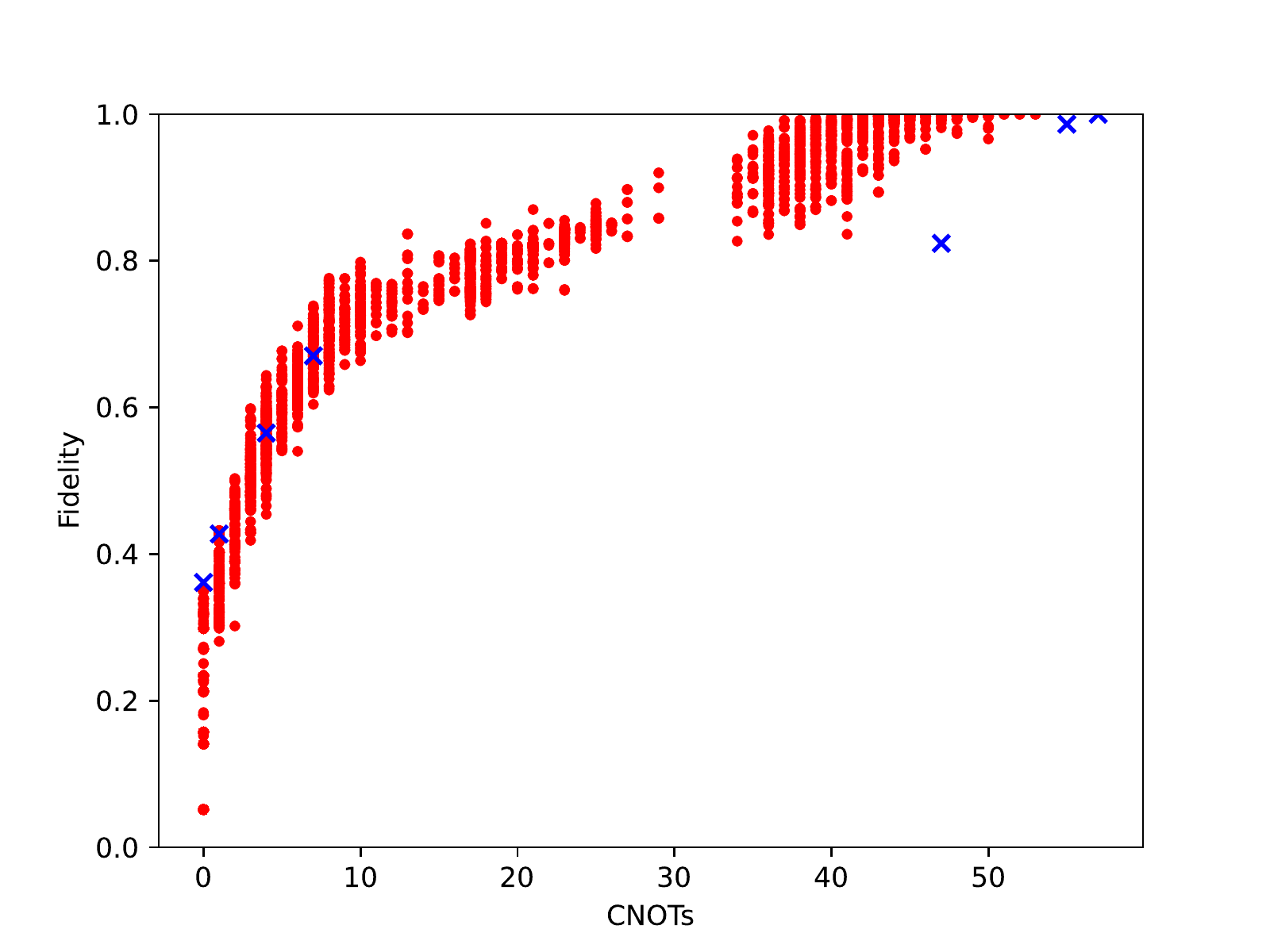}
	\includegraphics[width=0.5\textwidth]{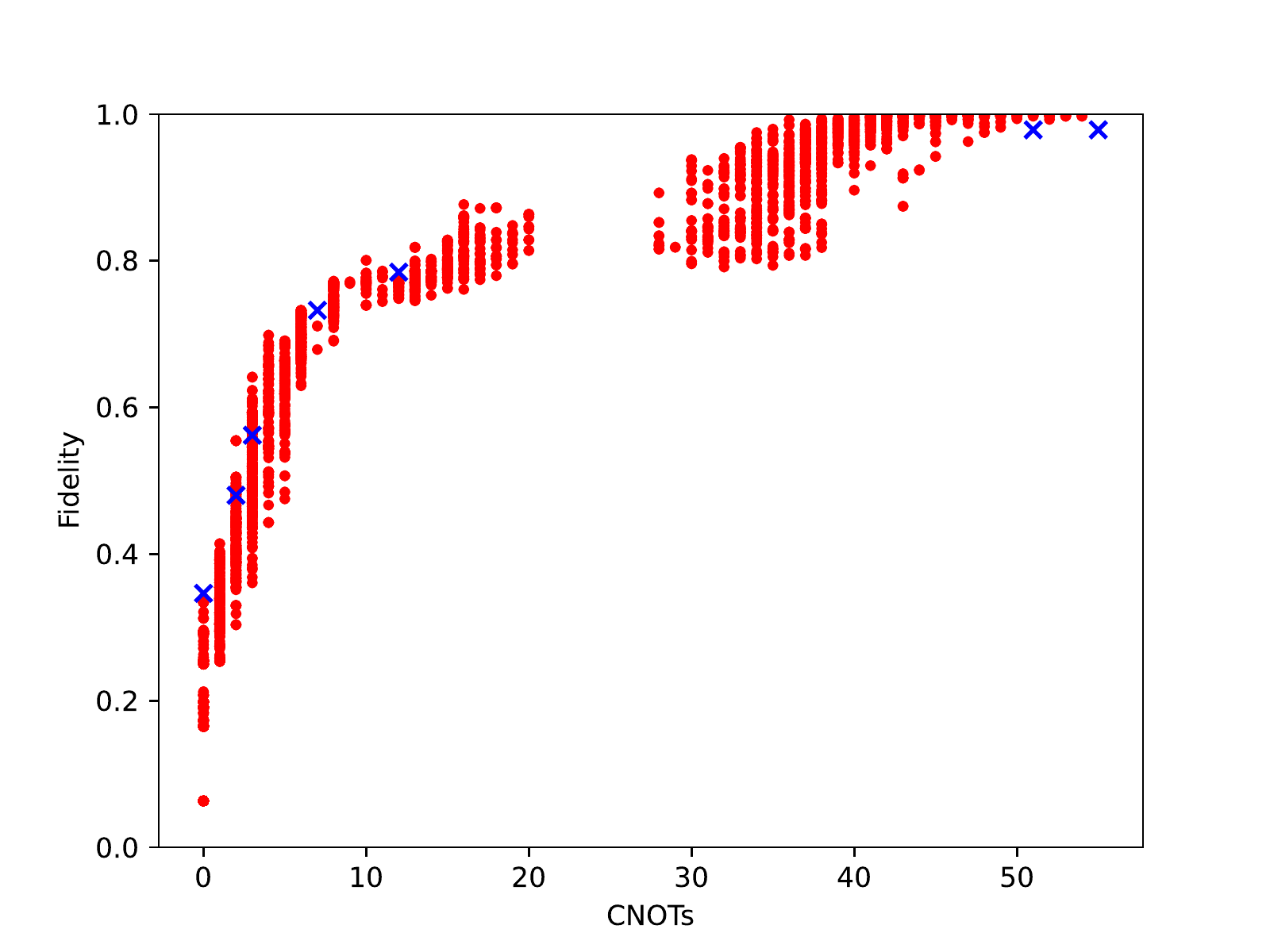}
	\caption{Example results for four arbitrarily picked random states (indices 160, 559, 748 and 932 in our database for the states in \cite{code}). Blue crosses: Circuits obtained by the LRSP algorithm. Red dots: Non-dominated circuits from 100 runs of the genetic algorithm for each state.}
	\label{fig:individualstates}
\end{figure}

The best non-dominated circuits found for a given state reproduce the approximately exponential convergence to the exact state as a function of the number of CNOT gates (in the absence of noise), as predicted by our theoretical analysis. See Figure \ref{fig:individualstates} for examples of results for four arbitrarily chosen individual random states. For each state the genetic algorithm was run 100 times, and the non-dominated solutions of the final populations are plotted in Figure \ref{fig:individualstates}. When we compare the circuits with the best noise-free fidelity for a given number of CNOT gates generated by the genetic algorithm and LRSP algorithm for a given state, we can see a significant rise in fidelity over LRSP results in most cases. This shows that the approximate state preparation circuits obtained from the LRSP algorithm are not near-optimal, at least when transpiled with the Qiskit transpiler for the specific quantum processor.

\begin{figure}[h!]
	\includegraphics[width=0.5\textwidth]{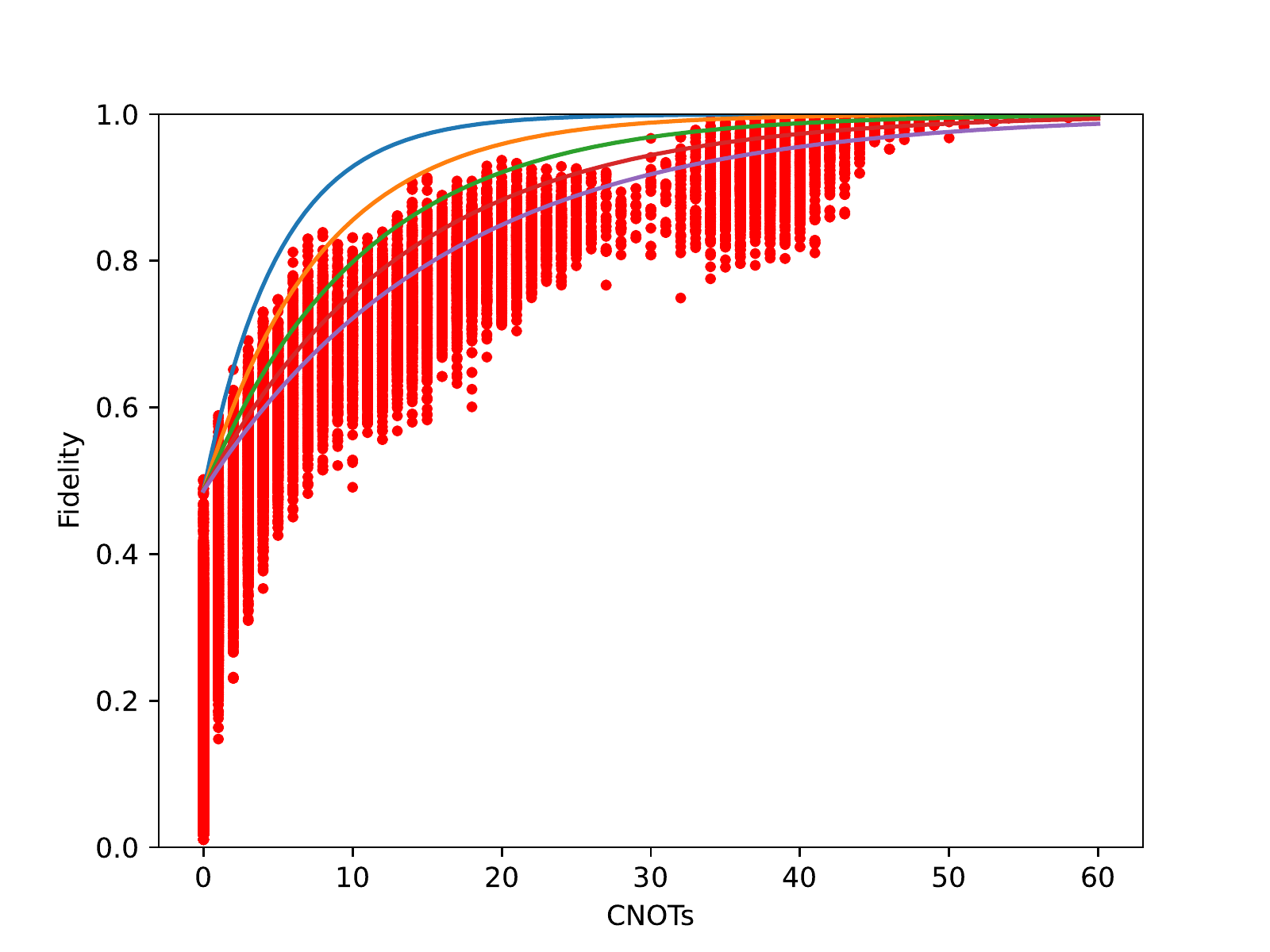}
	\includegraphics[width=0.5\textwidth]{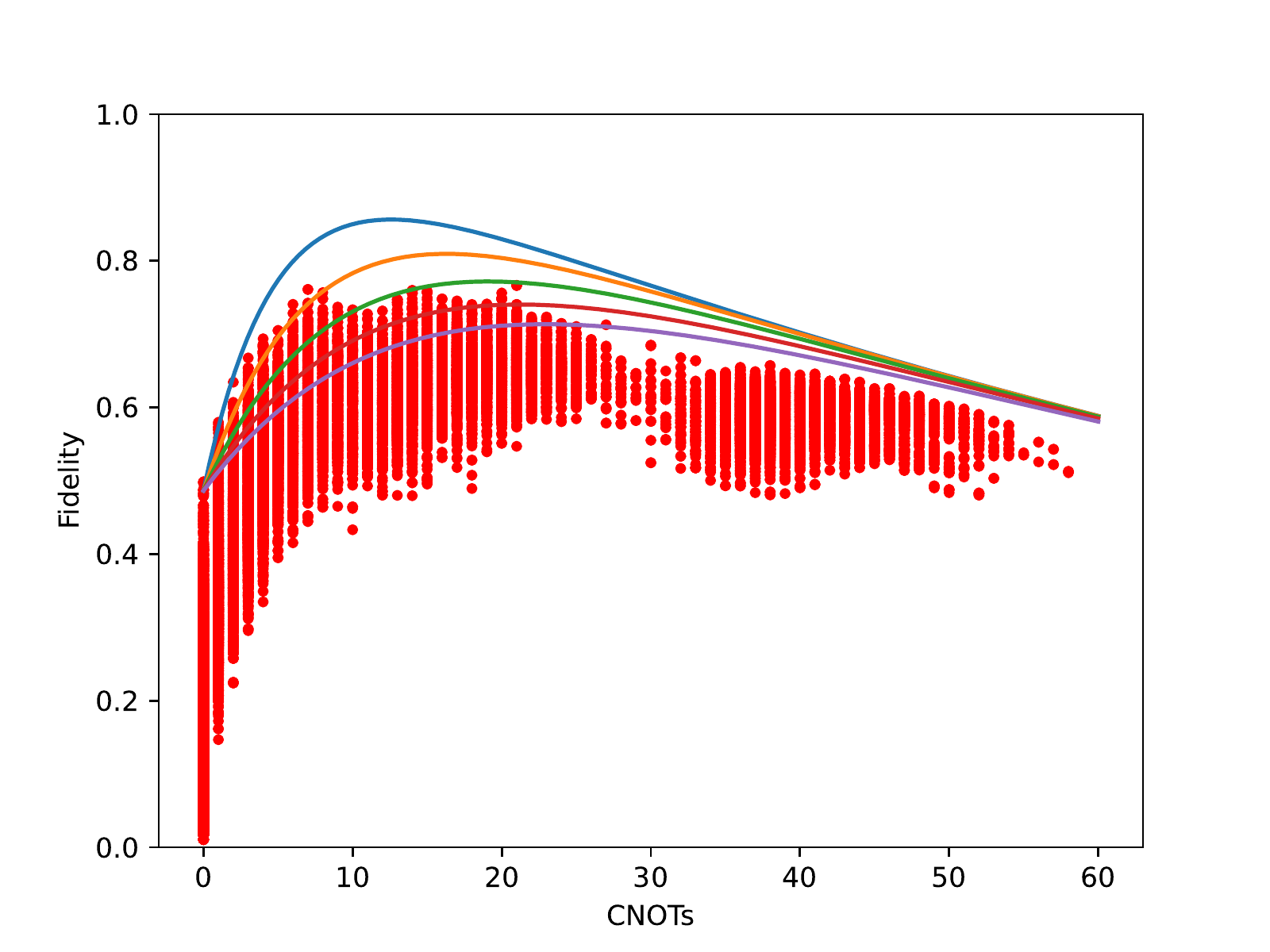}
	\caption{Numerical results and theoretical fidelity curves for 1000 random states without (left) and with (right) noise. Red dots: Non-dominated solutions of genetic algorithm runs for 1000 different random states. Parameters for the runs were 30 000 generations, population of 400, 5 qubits, and FakeVigo backend connectivity and noise model. Curves: Theoretical fidelity curves with $n=5$ and $p=0.0088$. $l_\text{ph}$ varies from $\langle d \rangle = 1.8$ (blue) to $1 + 6(\langle d \rangle - 1) = 5.8$ (purple) with unit steps.}
	\label{fig:theoreticalfit}
\end{figure}

For Figure \ref{fig:theoreticalfit}, we ran the genetic algorithm once for each one of 1000 random states, and plotted the non-dominated solutions of the final populations. Fitting of the theoretical fidelity curve derived in Subsection \ref{subsec:theory} to the data is also shown with $n=5$ and $l_\text{ph}$ ranging between the limiting values $\langle d \rangle = 1.8$ and $1 + 6(\langle d \rangle - 1) = 5.8$ with unit steps. The best fit with the optimal circuits is achieved with $l_\text{ph} \approx 3.8$ indicating that about this many physical CNOT gates are needed to implement one logical CNOT gate in the state preparation circuits, on average. The CNOT error rate of the backend is $p = 0.0088$. We see that the actual error rate seems to be somewhat higher, as the theoretical fit overestimates the noisy fidelities by a small margin. This is explained by the fact that also other errors not accounted for by our theoretical analysis, such as single-qubit gate errors and identity errors, contribute to the exponential decay of fidelity as a function of the circuit length.

\begin{figure}
	\includegraphics[width=0.5\textwidth]{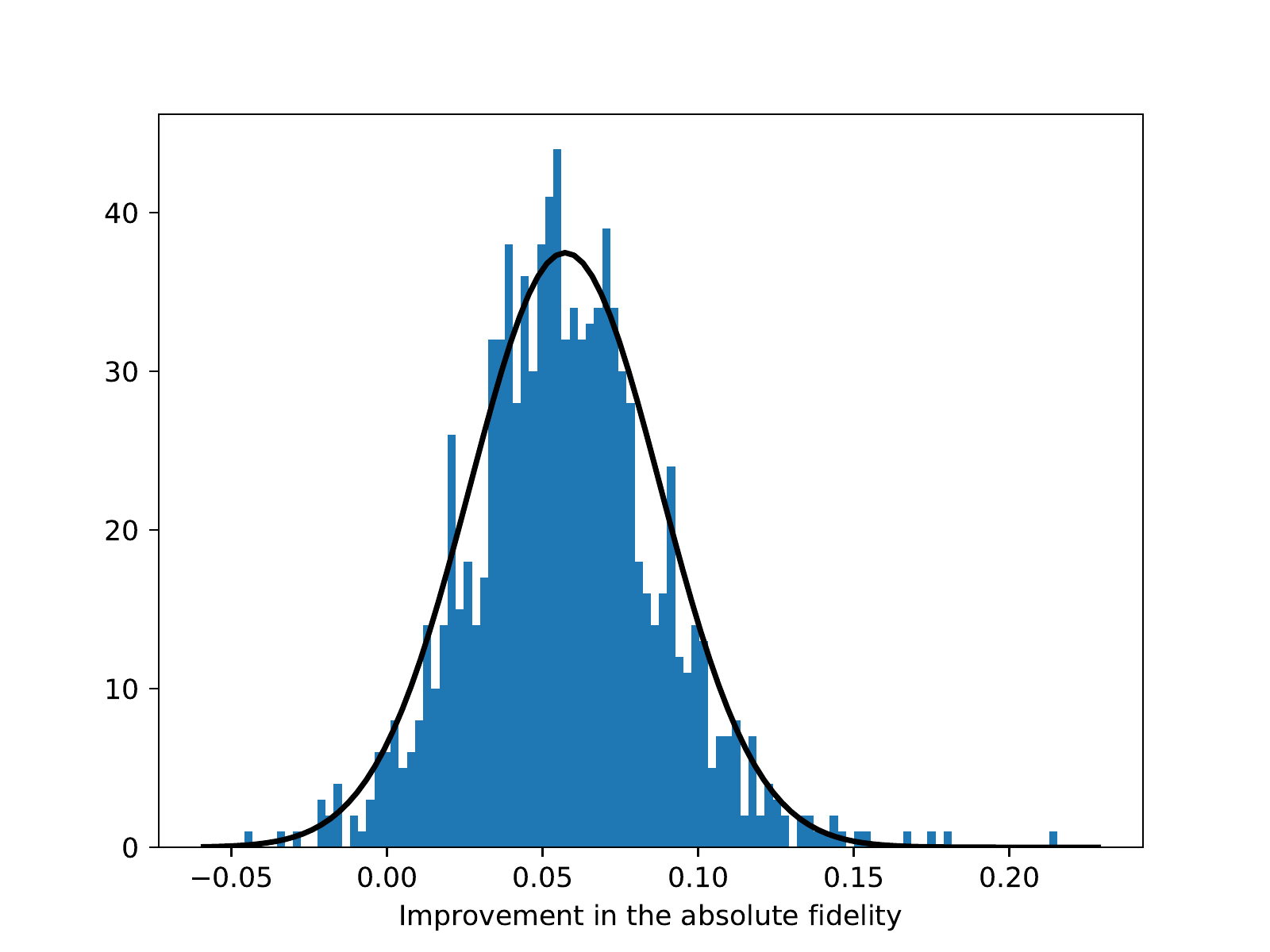}
	\includegraphics[width=0.5\textwidth]{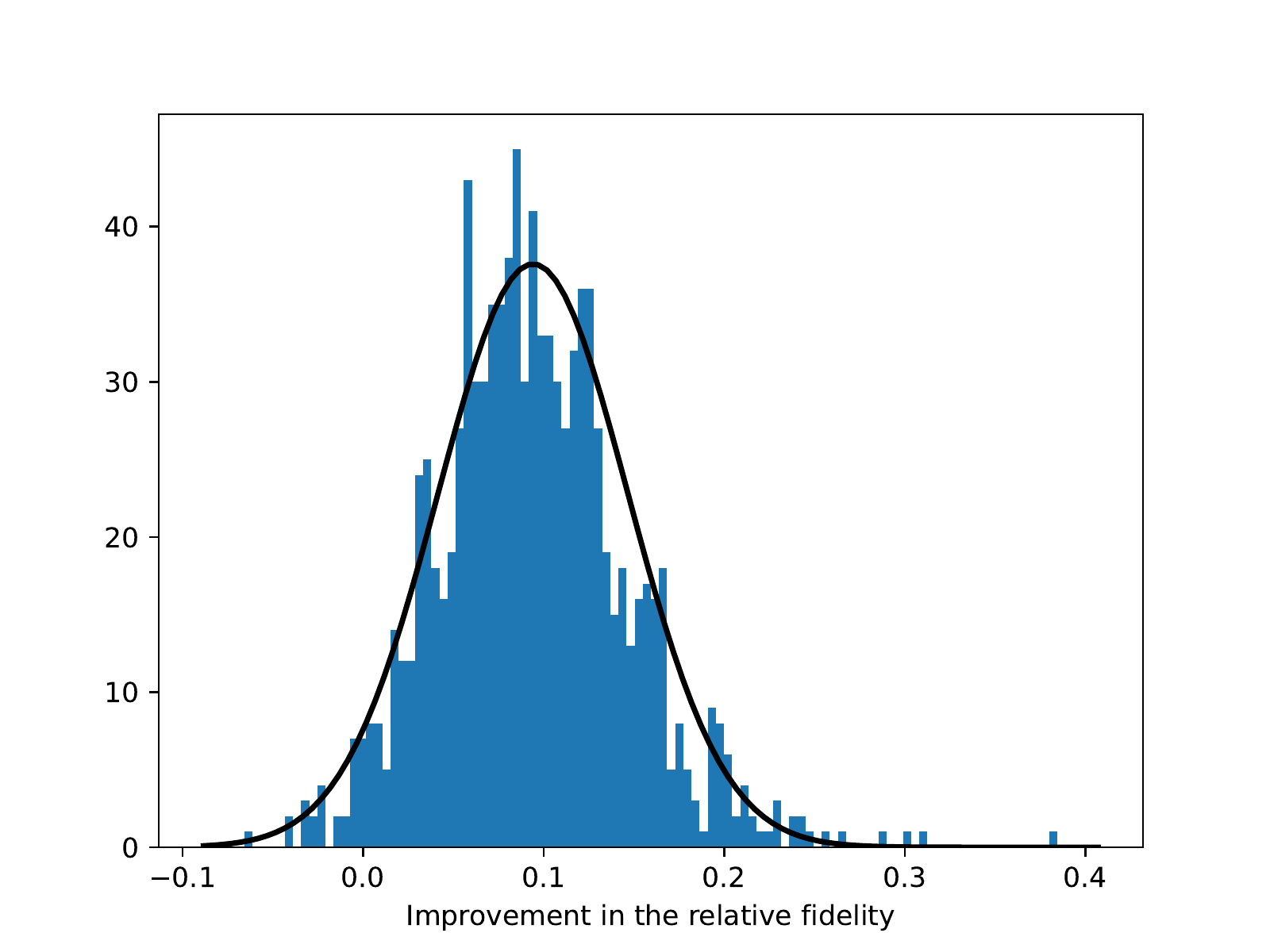}
	\caption{Left: The histogram of absolute differences in noisy fidelities between the genetic algorithm circuits and the LRSP circuits, together with a Gaussian fit. Right: The same for the relative differences in fidelity.}
	\label{fig:fid-ent}
\end{figure}

We may then consider including the noise in the simulation of the non-dominated solutions. Running the simulation with noise for the final populations shows that the highest total fidelity is achieved for shorter approximate circuits, as predicted by our theoretical analysis. For most of the runs improvements over LRSP circuits were achieved. Averaging over single genetic algorithm runs for 1000 random states gives an average absolute improvement of about 0.057 ($\sigma\approx 0.031$) in the maximum fidelity with noise, or an average relative improvement of about $9.4\%$ ($\sigma \approx 5.2\%$). The largest observed absolute improvement was about 0.22, and the largest relative improvement 38\%. In Figure \ref{fig:fid-ent} the distributions for the absolute and relative differences between the LRSP and genetic algorithm circuits are plotted. We see that in most cases the genetic algorithm is able to significantly improve on the circuits found by the LRSP algorithm. In a few cases the genetic algorithm was unable to improve on the performance of the LRSP circuit. In these cases it would be possible either to use the LRSP circuit directly, or to run the genetic algorithm again. It is likely that running the genetic algorithm more than once would improve the results, since (as we saw already in Figure \ref{fig:individualstates}) there is quite a lot of variation between the runs for an individual state.

\section{Summary \& outlook}\label{sec:summary}
We implemented a multi-objective genetic algorithm for quantum circuit discovery, and applied it to perform a numerical analysis of the approximate state preparation problem. We focused particularly on the native gate set,  connectivity and noise profile of the IBM 5-qubit processors. The genetic algorithm was able to improve on the circuits found by the low-rank state preparation (LRSP) algorithm introduced in \cite{Araujo21} and compiled with the Qiskit compiler in that it found preparation circuits with the same number of CNOT gates but a higher fidelity for the output state. As a result, also the maximum fidelity with realistic noise (implemented from the IBM Qiskit `FakeVigo' backend) was significantly improved in most cases with an average relative improvement of about 9.4\%, and maximum observed improvement of 38\%. We also provided a back-of-the-envelope theoretical analysis of the approximate state preparation problem to support our numerical findings, and derived the optimal fidelity curve, which was reproduced by our numerical results. Our results support the overall conclusions that (1) there is still room to improve in current state-of-the-art algorithms for approximate state preparation, (2) the maximum fidelity for generic state preparation on a NISQ computer may be achieved by using a shorter approximate circuit rather than the exact state preparation circuit, and (3) genetic algorithms continue to provide a useful tool for quantum circuit discovery and optimization, especially as the classical computing power at the disposal of researchers continues to increase.

In the future, our genetic algorithm can be applied to find near-optimal circuits for the state preparation task on NISQ computers. It would be interesting to search for optimized circuits for the preparation of some practically more relevant states than Haar random states, and see if the improvements can be as significant as for random states. Even though running the genetic algorithm is computationally expensive, it may still be beneficial to optimize the state preparation circuit to save the scarce quantum computational resources on a NISQ machine, especially if the particular state needs to be prepared many times during the desired algorithm. However, for a larger quantum processor the output state fidelity evaluation step will need to be implemented on the quantum machine itself, since the classical simulation becomes prohibitively expensive. Indeed, this seems to be a key challenge to overcome in quantum circuit compilation in general.

Finally, the idea of automated quantum circuit discovery is very appealing, and genetic algorithms seem to be well suited for the task. However, it would be great if we could somehow generalize the search from individual circuits, as done in this work, to families of circuits with different numbers of qubits, perhaps defined in some iterative manner. In this way it would be possible to search for general solutions to problems with different input sizes, as is often required. This might make it possible to numerically explore more directly the scaling of circuit depth as a function of the input size --- a key relation in determining the computational complexity of a problem --- and thus shed more light on the computational capabilities of quantum computers.

\section*{Acknowledgments}
We are grateful to Business Finland for supporting the `Post-Quantum Cryptography (PQC) Finland' project \cite{PQCFinland}, which partially funded this research. We acknowledge the computational resources provided by the Aalto Science-IT project, in particular, the Triton cluster. The authors also wish to acknowledge CSC – IT Center for Science, Finland, for computational resources.

\appendix

\section{Error estimate for approximate state preparation circuits}\label{app:error}
We may estimate the dependence of the noise-free error on the CNOT count by considering the fraction of the total volume of the Hilbert space, which can be reached with error $\epsilon$ by using at most $l$ CNOT-gates. The Hilbert space $\mathcal{H}_n$ of $n$ qubits is geometrically the $2(2^n-1)$-dimensional complex projective space $\mathbb{C}{\bf P}_{2^n-1}$. (See, e.g., \cite{Bengtsson17} for a review of complex projective geometry.) A complex projective space comes with a canonical metric, the Fubini-Study metric.\footnote{We normalize the metric so that the geodesic length $D_{FS}(|\phi \rangle , |\psi \rangle ) = \protect \qopname \relax o{arccos}(|\langle \phi |\psi\rangle |)$ between any two (normalized) state vectors $|\phi \rangle , |\psi\rangle \in \protect \mathbb {C}{\protect \bf P}_{d}$. With this normalization, any two orthogonal states have distance $\protect \frac {\pi}{2}$, and the unique closed geodesic passing through those states has circumference $\pi $.} The volume of a $2d$-dimensional complex projective space, as defined by this metric, is given by
\begin{align*}
	\text{Vol}(\mathbb{C}{\bf P}_{d}) = \frac{\pi^{d}}{d!} \,.
\end{align*}
Accordingly, the total volume of $\mathcal{H}_n$ is given by $\text{Vol}(\mathcal{H}_n) = \pi^{2^n-1}/(2^n-1)!$.

The subspace we reach from the initial state $|00\ldots 0\rangle$ without any CNOT gates is a tensor product $\mathcal{H}_1^{\otimes n}$ of single-qubit Hilbert spaces $\mathcal{H}_1$, each of which is isomorphic to a $2$-dimensional complex projective space of volume $\text{Vol}(\mathbb{C}{\bf P}_{1}) = \pi$. Thus, we get for the volume of the $2n$-dimensional subspace of states reachable by single-qubit unitaries $\text{Vol}(\mathcal{H}_1^{\otimes n}) = \pi^n$. Now, imagine `thickening' this subspace by distance $\epsilon$ into the total Hilbert space. For $\epsilon\ll 1$ we can approximate the thickened subspace as the Cartesian product $\mathbb{C}{\bf P}_{1}^{\times n} \times B_{d_\perp}(\epsilon)$, where $B_d(R)$ denotes the $d$-dimensional ball of radius $R$, and $d_\perp = 2(2^n-n-1)$ is the codimension of the subspace $\mathcal{H}_1^{\otimes n}$ in $\mathcal{H}_n$. The volume of a $d$-dimensional ball is given by
\begin{align*}
	\text{Vol}(B_d(R)) = \frac{\pi^{\frac{d}{2}}}{\Gamma(\frac{d}{2}+1)} R^d \,.
\end{align*}
Accordingly, the volume of the thickened subspace is given by
\begin{align*}
	\text{Vol}(\mathbb{C}{\bf P}_{1}^{\times n}\times B_{d_\perp}(\epsilon)) = \text{Vol}(\mathbb{C}{\bf P}_{1})^n \text{Vol}(B_{d_\perp}(\epsilon))) = \frac{\pi^{2^n-1}}{(2^n-n-1)!} \epsilon^{2(2^n-n-1)} \,.
\end{align*}

Without restrictions on the connectivity of the qubits, a single CNOT gate can be applied to any of the $\frac{n(n-1)}{2}$ pairs of qubits. Each allowed CNOT gate in a circuit increases the dimension of the subspace of states that can be reached without an error: After each CNOT gate two more single-qubit rotations can be added, which lead to distinct states. However, $z$-rotations of the control qubit and $x$-rotations of the target qubit commute with the CNOT gate, reducing the number of free parameters for each one of the rotations from 3 to 2. Each space of single-qubit rotations after a CNOT is isomorphic to a 2-sphere $S_2(1) \cong \mathbb{C}{\bf P}_1$, so we get an extra direct product factor of $\mathcal{H}_1^{2}$ for our subspace from each CNOT gate. Accordingly, with $l$ CNOT gates the volume of the subspace of states reached without an error is given by $\text{Vol}(\mathcal{H}_1^{\otimes (n + 2l)}) = \pi^{n+2l}$. Allowing for error $\epsilon$ now gives volume
\begin{align*}
	\text{Vol}(\mathbb{C}{\bf P}_{1}^{\times (n+2l)}\times B_{d_\perp}(\epsilon)) = \frac{\pi^{2^n-1} \epsilon^{2(2^n-(n+2l)-1)}}{(2^n-(n+2l)-1)!} \,.
\end{align*}

In the leading order, the number of different circuits containing $l$ CNOT gates is given by $\left( n(n-1)/2 \right)^l$. Some of these circuits are actually equivalent due to commuting CNOT gates and the saturation of the Hilbert space, but this only affects the sub-leading order behavior. Circuits with larger $l$ also reproduce the subspaces of states given by circuits with smaller $l$ as we may cancel out CNOT gates from the beginning of the circuit by setting the preceding single-qubit unitaries to identity, given that the initial state of the circuit is $|00\ldots0\rangle$. Assuming for simplicity that each different combination of CNOT gates leads to a distinct subspace of states\footnote{This assumption will presumably hold approximately true for $\epsilon \ll 1$, until we reach $\sim 2^n$ CNOT gates, which is required for the exact preparation of an arbitrary state.}, we reach a volume
\begin{align*}
	\text{Vol}(\mathbb{C}{\bf P}_{1}^{\times (n+2l)}\times B_{d_\perp}(\epsilon)) \left( \frac{n(n-1)}{2} \right)^l = \frac{\pi^{2^n-1} \epsilon^{2(2^n-(n+2l)-1)}}{(2^n-(n+2l)-1)!} \left( \frac{n(n-1)}{2} \right)^l
\end{align*}
with circuits containing at most $l$ CNOT gates. Dividing by the total volume of the Hilbert space, we get for the fraction of the volume covered by the subspace of states reached within an error $\epsilon$
\begin{align*}
	\eta_{n,l}(\epsilon) &= \frac{\text{Vol}(\mathbb{C}{\bf P}_{1}^{\times (n+2l)}\times B_{d_\perp}(\epsilon))}{\text{Vol}(\mathcal{H}_n)} \left( \frac{n(n-1)}{2} \right)^l \\
	&= \frac{(2^n-1)!\ \epsilon^{2(2^n-(n+2l)-1)}}{(2^n-(n+2l)-1)!} \left( \frac{n(n-1)}{2} \right)^l \,.
\end{align*}
Using the approximation $\Gamma(x+\alpha) \approx \Gamma(x)x^\alpha$ for $x\gg 1$, for large $n$ we may approximate
\begin{align*}
	\frac{(2^n-1)!}{(2^n-(n+2l)-1)!} \approx (2^n)^{n+2l} = 2^{n(n+2l)} \,.
\end{align*}
We then get
\begin{align*}
	\eta_{n,l}(\epsilon) &\approx 2^{n(n+2l)} \epsilon^{2(2^n-(n+2l)-1)} \left( \frac{n(n-1)}{2} \right)^l = 2^{n^2} \epsilon^{2(2^n-n-1)} \left(\frac{4^n}{\epsilon^4} \frac{n(n-1)}{2}\right)^l \,.
\end{align*}
We may then obtain at least a crude estimate for the error $\epsilon$, which is needed to cover the full Hilbert space by setting $\eta_{n,l}(\epsilon) = 1$ and solving for $\epsilon$. This gives
\begin{align*}
	\epsilon \approx \exp\left[ - \frac{(2 \ln 2)n + \ln\left( \frac{n(n-1)}{2} \right)}{2(2^n-n-1)} l - \frac{(\ln 2)n^2}{2(2^n-n-1)} \right] \,.
\end{align*}

In the above analysis we did not yet consider the possible restrictions on the connectivity of qubits. We may estimate the effect of restricted qubit connectivity by dividing $l$, the number of logical CNOT gates in the circuit, by the average number $l_\text{ph}$ of physical CNOT gates required to implement one logical CNOT gate. Thus, we arrive at an estimate for the maximum error with $l$ CNOT gates:
\begin{align*}
	\epsilon \approx \exp\left[ - \frac{(2 \ln 2)n + \ln\left( \frac{n(n-1)}{2} \right)}{2(2^n-n-1)} \frac{l}{l_\text{ph}} - \frac{(\ln 2)n^2}{2(2^n-n-1)} \right] \,.
\end{align*}
A linear scaling of the number of CNOT gates is obviously a very simplistic way to take the connectivity into account. However, we know that different connectivities differ only polynomially, so the overall exponential behavior of $\epsilon$ must be preserved. We also get a decent fit to the numerical results using this approach, as we see in Section \ref{sec:results}.

\section{Asymptotic analysis of the fidelity formula}\label{app:asymptotics}

Let us consider the asymptotic scaling of $l_*$ according to the formula
\begin{align*}
	\frac{l_*}{l_\text{ph}}  = \frac{1}{A(n)} \ln\left( 1 - \frac{A(n)}{\ln(1-p)} \right) - \frac{B(n)}{A(n)}\,,
\end{align*}
where
\begin{align*}
	A(n) = \frac{(2 \ln 2)n + \ln\left( \frac{n(n-1)}{2} \right)}{2^n-n-1} \,,\quad B(n) = \frac{(\ln 2)n^2}{2^n-n-1}
\end{align*}
as we increase the number of qubits $n$. We have
\begin{align*}
	x \ln(1-\frac{1}{cx}) \rightarrow -\frac{1}{c}
\end{align*}
as $x\rightarrow \infty$, so the first term converges to the constant $\left(\ln\frac{1}{1-p}\right)^{-1}$ as $n\rightarrow\infty$, since $A(n) \rightarrow 0$ in the limit. The second term
\begin{align*}
	\frac{B(n)}{A(n)} = \frac{(\ln 2)n^2}{(2 \ln 2)n + \ln\left( \frac{n(n-1)}{2} \right)} \approx \frac{n}{2}
\end{align*}	
when $n\gg 1$. Thus, for large $n$, we have the approximation
\begin{align*}
	\frac{l_*}{l_\text{ph}}  \approx \left(\ln\frac{1}{1-p}\right)^{-1} - \frac{n}{2}\,.
\end{align*}
First of all, to get any benefit from the computation and reach a better average fidelity than with the initial state $|00\ldots0\rangle$, we must have $l_*>0$, which gives the condition $p < 1 - e^{-2/n} \approx \frac{2}{n}$ for $n\gg 1$. Thus, the error rate of CNOT gates must decrease inversely proportionally to the number of qubits to have any chance of preparing an arbitrary target state to any degree of approximation. On the other hand, counter-intuitively the optimum length $l_*$ of a state preparation circuit actually \emph{decreases} linearly at fixed error rate as the number of qubits is increased. This occurs because it gets exponentially hard to get close to an arbitrary state. It doesn't pay off to try that hard with noisy gates, because the errors build up as we try to increase the length of the circuit to get closer to the target state. Thus, we are better off with a shorter circuit, which at least gives somewhat a decent approximation, rather than trying to climb up the exponentially long ladder with faulty rungs towards the target state.

When $p\ll 1$, we may further approximate $\ln\frac{1}{1-p} \approx p$. Assuming again that $n\gg 1$, the maximum value of total fidelity is found to be approximated by
\begin{align*}
	F(\rho_\text{noise}, |\text{target}\rangle )_\text{max} &\approx (1-p)^{l_*} \left(1 - \exp\left[  - A(n) \frac{l_*}{l_\text{ph}} - B(n) \right] \right) \\
	&\approx (1-p)^{\left(\frac{1}{p} - \frac{n}{2}\right)l_\text{ph}} \left( 1 - \frac{2^n\ln(1-p)}{(2 \ln 2)n} \right)^{-1} \,.
\end{align*}
In order to keep the maximum total fidelity sufficiently high, both of the factors in this expression must remain large enough as $n$ grows. In order to keep the first term constant, $(1-p)^{\left(\frac{1}{p} - \frac{n}{2}\right)l_\text{ph}} = f_1$, we find that the error rate must satisfy $p \approx 2(1 + \frac{\ln f_1}{l_\text{ph}})/n$, which again implies inverse proportionality of the error rate to the number of qubits. However, the second factor imposes a stricter requirement. In order for it to stay constant,
\begin{align*}
	\left( 1 - \frac{2^n\ln(1-p)}{(2 \ln 2)n} \right)^{-1} = f_2 \,,
\end{align*}
the error rate must satisfy
\begin{align*}
	p \approx (\ln 2)(f_2^{-1} - 1) 2^{-n + \log_2 n} \,.
\end{align*}
We find a nearly inverse exponential scaling requirement for the error rate in the number of qubits, which reflects the general difficulty of the arbitrary state preparation task in the presence of noise.

\end{document}